\documentclass[a4paper]{article}

\usepackage{icrc2013}

\title{At what Rigidity does the Solar Modulation of Galactic Cosmic Rays begin?}

\shorttitle{Rigidity where modulation begins}

\authors{M.S. Potgieter and R. du T. Strauss 
}

\afiliations{
Centre for Space Research, North-West University, Potchefstroom, South Africa\\ 
}

\email{Marius.Potgieter@nwu.ac.za}

\abstract{Observationally, it is difficult to establish at what rigidity the modulation of galactic cosmic rays (CRs) actually begins in the heliosphere. 
It should be possible to do if the relevant local interstellar CR spectra were known and reliable measurements were made between 10 GV and $\sim$200 GV, inside and outside the heliosphere. 
Numerical models for solar modulation studies are based on simply assuming that CR modulation begins at a given spatial boundary and at rigidities between 30-50 GV, usually handled as an initial condition. 
The Stochastic Differential Equations approach to numerical modelling presents the opportunity to determine the level of modulation from high to low rigidities because an initial condition at a prescribed high rigidity is not required. 
We present the results of such an approach and show the percentage modulation of CR protons as a function of kinetic energy between 100 MeV and 250 GeV. }


\begin{document}
\maketitle

\section{Introduction}

It is difficult to establish at what rigidity the modulation of galactic cosmic rays (CRs, e.g. protons) actually begins in the heliosphere. This could change as a function of time and position. 
Observationally, this could be done if the relevant local interstellar spectra (LIS) were precisely known at appropriate rigidities and reliable measurements are made simultaneously inside and outside the heliosphere over a relatively large rigidity range, e.g. between 10 GV and $\sim$100 GV. 
Intensifying this challenge is the controversial issue of where exactly does the solar modulation of CRs begin in space (position). 
Current knowledge stipulates that modulation begins at the heliopause, the place where the solar wind and heliospheric magnetic field subside to have negligible influence on CRs. 
Such a region was observed in August 2012 by the spacecraft Voyager 1 at about 122 AU from the Sun [1]. 
It is still unclear if this is the 'genuine' heliopause as predicted by MHD models because the magnetic field data from Voyager 1 did not indicate a transition to an interstellar magnetic field. 

Recent numerical modelling indicates that the modulation of CRs should commence where the heliosphere first begins to disturb the local interstellar medium, such as from the bow wave in the nose direction of the heliosphere [2]. 
Where modulation spatially begins in the tail direction is much less certain. However, the total modulation of CRs inside the outer heliosheath 
(region beyond the heliopause up to the bow wave) is much less than inside the heliopause, upstream towards the Sun. How large the total effect is, depends on the energy of the CRs 
and how disturb the modulation region is in terms of the turbulence that occurs. This modulation region, inwards towards Earth, typically consists of an outer heliosheath, 
of which we know very little, then the inner heliosheath, which had been explored by Voyager 1 and still is by Voyager 2, the solar wind termination shock, the outer and inner heliosphere.  

Numerical models for the solar modulation of CRs can of course be utilized to try and find an answer to the problem as portrayed. 
But, also this is not straight forward. All 'standard' numerical models are based on simply assuming that CR modulation begins at a given modulation boundary and at rigidities between 30-50 GV. 
The latter is usually handled as an initial condition, for example, within the well-known Alternating Direction Implicit (ADI) numerical scheme (e.g. [3,4]). 

Over the past few years, the Stochastic Differential Equations (SDE) approach to numerical modelling has become popular, increasingly so also for solar modulation studies (e.g. [5,6]). 
The reason is that the numerical scheme and process are perfectly parallelizable and is therefore very sufficient in utilizing the advantage given by large computer clusters. 
This approach offers, from a physics point of view, the opportunity to determine the level of modulation for any given rigidity (or kinetic energy/nucleon), 
from very high to low rigidities because an initial condition is not required. 

The results of such an approach are presented, showing the modulation fraction of galactic protons at a given kinetic energy, between the modulation boundary and Earth.

\begin{figure}[!ht]
 \centering
 \includegraphics[width=0.52\textwidth]{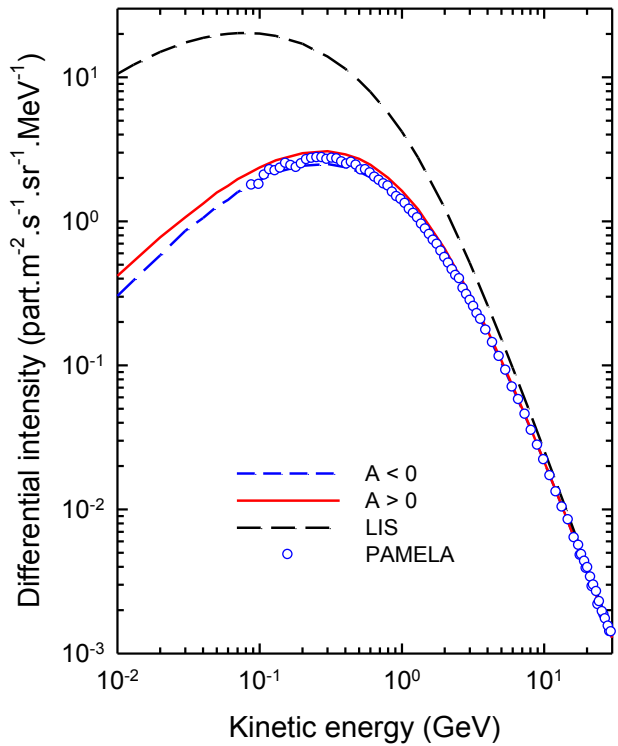}
 \caption{Computed modulated proton spectra at Earth for the two drift cycles (A $>$ 0 and A $<$ 0) during solar minimum activity conditions with respect to the local interstellar spectrum (LIS), 
and in comparison with the PAMELA proton observations at the end of 2009 [10]. The present cycle is an A $<$ 0 cycle. Note that on this scale, it appears if no modulation occurs above $\sim$20 GeV.}
\label{fig:Figure1}
\end{figure}

\section{Modelling approach}

The SDE approach and subsequent numerical model for this CR modulation study is based on the model described by Strauss et al. [6]. 
For additional information on the approach and how it is developed and bench marked, see also [2,6] and for a generalized description, see [5].  

The approach allows that one can trace single pseudo-particles, with any given energy, from the heliospheric modulation boundary to Earth
(or in the opposite direction if the time-backwards approach is followed). In principle, this means that one could  'count' these 'particles' to establish 
how many of those that have entered the heliosphere (crossing the modulation boundary) can actually arriving at Earth. 
This must of course then be repeated for thousands of 'particles' to establish a reasonable statistical significance and be repeated many times to obtain a trustworthy CR spectrum, 
which could be a quite tedious process.

This approach and procedure can be seen as a numerical experiment, comparing the intensity for a given energy at the modulation boundary to the modulated values at Earth, 
doing in fact what is hard to achieve observationally. This ratio is presented as the modulation fraction. The modulation boundary for this study was specified at 130 AU. 
The model was run to determine the level of modulation between the modulation boundary and Earth from 100 MeV to 250 GeV for galactic protons.

The level of modulation anywhere in the heliosphere is determined also by the rigidity dependence of the three major diffusion coefficients and that of the drift coefficient [3,4,6,7,8]. 
For example, if the diffusion coefficients, such as parallel and perpendicular to the mean magnetic field, change simply proportional to rigidity above 5 GV (that is $P^{1.0}$) 
the modulation will be significantly more than when they scale like $P^{2.0}$ [see illustrations from e.g. 6,11,12]. Keep in mind that any diffusion coefficient also scales with the velocity of the particles.
For this study we scaled the parallel diffusion coefficient 
proportional to $P^{2.0}$ with $P > 10$ GV but the perpendicular diffusion coefficient proportional to $P^{1.67}$. The first mentioned dominates in the inner heliospheric equatorial regions
whereas the latter dominates in the outer heliosphere. At lower rigidities both scale proportional to $P^{0.3}$, with the perpendicular diffusion coefficient much smaller than the parallel diffusion coefficient;
see [16] for an elaborate discussion of the differences between the three major diffusion coefficients. For additional applications to other aspects of the solar modulation of CRs, see [2,6,9,12,15].
Furthermore, the drift coefficient also scales with $P^{1.0}$ with $P > 1$ GV, as long as weak scattering is assumed; see also [11,12,15,16].  

The solar wind velocity was assumed to be directed radially outwards, with a latitude dependent speed during solar minimum conditions.
The heliopause, where the local interstellar spectrum (LIS) is specified, was assumed to be located at 130 AU. For protons the LIS of Langner et al. [14] was used. 
A constant heliospheric current sheet (HCS) tilt angle of $\alpha=5^{\circ}$ was assumed, 
while an unmodified solar magnetic field was implemented, normalized to a magnitude of 4 nT at Earth.These values represent ideal solar minimum conditions,

All relevant parameters and assumptions as described by [6,7] are not repeated here so it suffices to mention that all four 
major modulation mechanisms are applied, including particle drifts, in solving the transport equation [13]. 
Because of drifts, the modulation study was repeated for the so-called A $>$ 0 (e.g. around 1976 and 1997) 
and A $<$ 0 (e.g. around 1987 and 2009) polarity cycles of the solar magnetic field. For a review on drifts and charge-sign dependent modulation and other processes, see [8,14]. 

\begin{figure}[!t]
\centering
			\includegraphics[width=0.51\textwidth]{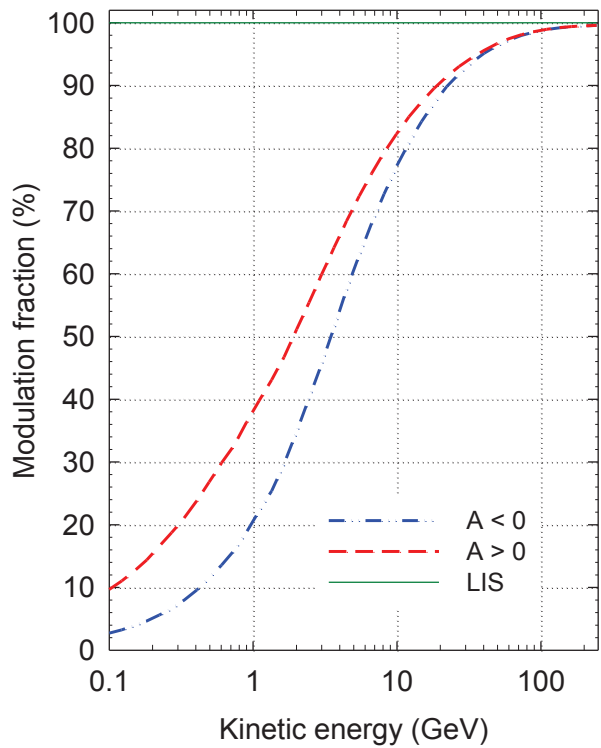}
			\caption{Modulation fraction as the ratio between the LIS intensity (100$\%$) at the modulation boundary and the two modulated spectra at Earth for the two drift cycles as in figure 1. 
			Solar minimum modulation conditions with ideal (full) drift conditions were assumed.}
\label{fig:Figure2}
\end{figure}

\section{Results and Discussion}

Figure 1 shows three spectra with differential intensity as a function of kinetic energy for galactic protons. The highest one is what was used as input spectrum, that is, 
the LIS, assumingly the same as the galactic spectrum. 
The other two are the modulated spectra for solar minimum activity conditions at Earth for the two mentioned drift cycles. 
A major modulation feature caused by particle drifts is that the A $<$ 0 cycles are always higher than the A $<$ 0 cycles at energies below a few GeV [e.g. 9,11]. 
In the figure the A $<$ 0 spectrum is compared to the observed spectrum for the end of 2009 from the PAMELA experiment [10].  This illustrates that the level of modulation (and the relevant modulation parameters) as modelled, is realistic. 

Figure 2 shows the modulation fraction, in percentage, as a function of kinetic energy. This is the ratio between the proton LIS intensity (100$\%$) 
at the modulation boundary and the two modulated spectra at Earth as given in figure 1.  
This shows that for both drift cycles there is a 0.5$\%$ reduction at 200 GeV in intensity between the modulation boundary and Earth. 
This fraction is given in Table \ref{table1} for decreasing energy and for both drift cycles. 

\begin{table}[h]
\begin{center}
\begin{tabular}{|l|c|c|}
\hline Kinetic Energy & A $>$ 0 cycle & A $<$ 0 cycle \\ \hline
200 GeV	& 0.4$\%$	& 0.4$\%$  \\ \hline
110 GeV	& 1.0$\%$	& 1.1$\%$  \\ \hline
60 GeV	& 2.6$\%$	& 2.8$\%$ \\ \hline
40 GeV	& 4.4$\%$ &	4.9$\%$ \\ \hline
22 GeV	& 8.6$\%$	&10.2$\%$ \\ \hline
10 GeV	& 17.7$\%$ & 22.9$\%$ \\ \hline
3 GeV	& 40.1$\%$ &	54.3$\%$ \\ \hline
1 GeV	& 61.2$\%$ & 79.1$\%$ \\ \hline
500 MeV	& 73.2$\%$	& 88.5$\%$ \\ \hline
100 MeV	& 90.3$\%$	& 97.3$\%$ \\ \hline
\end{tabular}
\caption{Modulation level (specified as 100$\%$ minus the modulation fraction) between Earth at 1 AU and the heliospheric modulation boundary at 130 AU for selected CR proton energies and for both drift modulation cycles during ideal solar minimum conditions.}
\label{table1}
\end{center}
\end{table}

The effect of drifts is small above 40 GeV, but below this energy the difference between A $>$ 0 and A $<$ 0 intensities 
becomes progressively larger as the energy decreases; at 10 GeV this increases to a 5$\%$ difference, 
with a maximum difference of about 17.5$\%$ around 1.5 GeV. Below this energy the difference becomes 
somewhat smaller but maintains almost the same value as the adiabatic energy losses become the dominate process.
If no drifts would be assumed and simulated, the difference between the two polarity cycles would disappear, of course, and the modulation fraction would be different, closer to the A $>$ 0 solutions than to the A $<$ 0 values (see figure 2, also Table 1).
These values will certainly become larger if solar maximum conditions are simulated. The drift effect may then subside [8,17].

It should be noticed that there were solar minimum periods when the consecutive proton spectra crossed at a given rigidity to cause the A $>$ 0 spectrum to be higher than the A $<$ 0 below a few GeV but not at higher energies [11, see their figure 2]. 

\section{Conclusion}

Assuming that the solar modulation of galactic protons begins at the heliopause, taken here at 130 AU, the kinetic energy where the modulation fraction reaches 1$\%$ is found to be at 110 GeV. 
The outcomes (see Table \ref{table1}) indicate that solar modulation affects CRs up to 200 GeV to reach a reduction in intensity of 5 $\%$ around 40 GeV, 10$\%$ around 20 GeV and 20$\%$ around 10 GeV. 
Assuming solar modulation to commence at 50 GeV, as is done for many modulation studies, seems justified.

These results should be seen in the context of the assumptions made for the modulation parameters because the modulation fraction can 
be increased or decreased depending on what is assumed for the rigidity dependence of the three major diffusion coefficients and the drift coefficient over the relevant energy range. 

What is shown in Table 1, represents ideal solar minimum conditions, so that these values will certainly become larger if solar maximum conditions are simulated. The drift effect may then subside.

\vspace*{0.5cm}
\footnotesize{{\bf Acknowledgement:}{~The partial financial support of the South African National Research Foundation (NRF), 
the SA High Performance Computing Centre (CHPC) is acknowledged.}}

\end{document}